\documentstyle[preprint,aps]{revtex}

\tightenlines

\begin{document}

\draft

\title{\bf Various series expansions for\\
the bilayer $S={1\over 2}$ Heisenberg antiferromagnet}
\author{Zheng Weihong\cite{byline1}} 
\address{School of Physics,                                              
The University of New South Wales,                                   
Sydney, NSW 2052, Australia.}                      


\maketitle 

\begin{abstract}
Various series expansions have been developed  for the
two-layer, $S={1\over 2}$, square lattice Heisenberg antiferromagnet.
High temperature expansions are used to calculate the
temperature dependence of the susceptibility and specific heat.
At $T=0$, Ising expansions are used to study the properties of the 
N\'{e}el-ordered phase, while dimer expansions are used
to calculate the ground-state properties and excitation spectra of
the magnetically disordered phase.
The antiferromagnetic order-disorder transition point is determined to
be $(J_2/J_1)_c=2.537(5)$. Quantities computed include 
the staggered magnetization, the  susceptibility, the triplet
spin-wave excitation spectra, the spin-wave velocity, 
and the spin-wave stiffness. We also estimates that the ratio of
the intra- and inter-layer exchange constants to be
$J_2/J_1\simeq 0.07$ for cuprate superconductor 
YBa$_2$Cu$_3$O$_{6.2}$.
\end{abstract}                                                              


\narrowtext
\section{INTRODUCTION}
The magnetic properties of low dimensional systems have been  the subject
of intense theoretical and experimental research in recent years.
It is by now well established that one-dimensional Heisenberg antiferromagnets
with integer spin have a gap in the excitation spectrum, whereas 
those with half-integer spin  have gapless excitations.
The former have a finite correlation length, while
for the latter it is infinite with the spin-spin
correlation function decaying to zero as a power law.
In 2-dimensions, the unfrustrated square-lattice Heisenberg model
has long range N\'{e}el order in the ground state. It has gapless
Goldstone modes as expected.
In recent years much interest has focussed on systems
with intermediate dimensionality and on questions of crossovers
between $d=1$ and $d=2$, and the investigations showed
that the crossover from the single Heisenberg chain to the
two-dimensional antiferromagnet  square lattice, obtained
by assembling chains to form a ``ladder'' of increasing width,
is far from smooth. Heisenberg ladders with an even
number of legs (chain) show a completely different behavior
than odd-leg-ladders. While even-leg-ladders have a spin gap
and short range correlation, odd-leg-ladders have no
gap and power-law correlations.

This paper is concerned with a system between 2 and 3 dimensions:
the $S={1\over 2}$ two-layer Heisenberg antiferromagnet, where
each layer is composed of a nearest-neighbor Heisenberg model on a square
lattice, and there is a further antiferromagnetic coupling 
between corresponding sites of each layer. The system can be described
by the following Hamiltonian:
\begin{equation}
H = J_1 \sum_{\alpha=1,2} \sum_{\langle i,j\rangle} 
{\bf S}_{\alpha,i} \cdot {\bf S}_{\alpha,j} + 
J_2 \sum_i {\bf S}_{1,i} \cdot {\bf S}_{2,i} \label{H}
\end{equation}
where ${\bf S}_{\alpha,i}$ is a S=1/2 spin operator at 
site $i$ of the $\alpha$th layer, and 
$\langle i,j\rangle$ denotes a pair of nearest neighbor sites on
a square lattice. Here $J_1$ is the interaction between 
nearest neighbor spins in one plane, 
and $J_2$ is the interaction between
adjacent spins from different layers.
We denote the ratio of couplings as $y$,
that is, $y=J_2/J_1$.
In the present paper, we study only the case of antiferromagnetic coupling, 
that is, both $J_1$ and $J_2$ are positive, although some of the 
series are applicable to ferromagnetic coupling case.
In the small $J_2/J_1$ limit,
the model describes two weakly interacting 2D Heisenberg antiferromagnets,
where each of them is  N\'{e}el ordered and possesses gapless Goldstone
excitations. 
In the large $J_2/J_1$ limit, pairs of adjacent spins from 
different layers form spin singlets separated from 
triplet states by a gap, $\sim J_2$, 
and the system has no long-range order. Thus one expects there
is a transition from ordered antiferromagnet to a quantum disordered spin
liquid at a certain critical ratio of $J_2/J_1$.
The bilayer Heisenberg antiferromagnets has recently attracted
attention for several reasons. 
On the experimental side, 
the model is relevant
to the understanding  of the magnetic properties of 
cuprate superconductors YBa$_2$Cu$_3$O$_{6+x}$
containing two  weakly coupled CuO$_2$ layers\cite{rez,hay,mil96}. 
On the theoretical side,
it is probably the simplest unfrustrated spin system which displays a
quantum disordering transition of $O(3)$ universality class, and Quantum 
Monte Carlo (QMC) simulation on this model do not suffer from a sign
problem, making it possible to obtain precise data on the
critical properties of the quantum transition.
Hida\cite{hid} has performed a dimer expansion up to 6th order
in $J_1/J_2$, and
he found the critical ratio to be $(J_2/J_1)_c =2.56$,
Gelfand\cite{gel} has recently extended the dimer expansion to 8th order,
and obtains the critical ratio $(J_2/J_1)_c = 2.54(2)$.
Sandvik {\it et al.}\cite{san94,san95,san96} have carried out a detailed QMC
simulations for the finite-temperature properties
of small $(J_2/J_1)$ and near-critical $(J_2/J_1)_c$ systems, 
and found a similar critical ratio:  $(J_2/J_1)_c =2.51(2)$.
On the other hand, a Schwinger boson mean-field calculation
by Millis and Monien\cite{mil} resulted in a very large critical
ratio:  $(J_2/J_1)_c =4.48$;
a self-consistent spin-wave theory\cite{chu95} also predicts a large
critical ratio: $(J_2/J_1)_c =4.3$. 
Chubukov and Morr\cite{chu95} have argued
that the discrepancies between the spin-wave results and the
numerical simulations (or series expansions) have a physical
origin and are related to the fact that in the spin-wave approach one 
neglects longitudinal spin fluctuations, and after taking 
longitudinal spin fluctuations into account, they got the critical
ratio: $(J_2/J_1)_c =2.73$, which is reasonably close to
the numerical results. The critical ratio obtained
by the Schwinger-boson Gutzwiller-projection method\cite{miy}
is $(J_2/J_1)_c =3.51$.
 
We have carried out extensive series studies of this system, 
via Ising expansions and dimer expansions at $T=0$, and
also by high temperature series expansions. 
Our results confirm the existence of a 
order-disorder transition at critical coupling ratio
$J_2/J_1 = 2.537(5)$.
The complete spin-wave excitation spectra, spin-wave velocity, and some
order parameters are computed. 
We also compare our results with previous calculations.

This paper is organized as follows: In Sec. II the dimer 
expansions are presented, and the antiferromagnetic 
order-disorder transition
points is determined precisely.
In Sec. III, the Heisenberg-Ising model is studied via 
the Ising expansions, which allow for estimates of the 
domain stability of Ne\'el-order phase, and also allow a 
direct comparison with experimental data of  
YBa$_2$Cu$_3$O$_{6+x}$. In Sec. IV we present
the results of the high-temperature expansions, 
and our conclusion is presented in the last section.

\section{Dimer expansions}
In the limit that the exchange coupling along the rungs $J_2$
is much larger than the coupling $J_1$ within the plane, 
that is $y\gg 1$, the rungs interact only weakly
with each other, and the dominant configuration 
in the ground state is the product state with the spin 
on each rung forming a spin singlet. The Hamiltonian in Eq. (\ref{H})
can be rewritten as,
\begin{equation}
H/J_2 =  H_0 + (1/y) V  \label{Hdimmer}
\end{equation}
where
\begin{eqnarray}
H_0 &=& \sum_{i} {\bf S}_{1,i} \cdot {\bf S}_{2,i} \nonumber \\  && \\
V &=& \sum_{\alpha=1,2} \sum_{\langle i,j\rangle} 
{\bf S}_{\alpha,i} \cdot {\bf S}_{\alpha,j}  \nonumber 
\end{eqnarray}
We can construct expansion in $1/y$ by treating
the operator $H_0$ as the unperturbed Hamiltonian
and the operator $V$ is treated as  a perturbation.
The linked-cluster expansion method 
has been previously reviewed in several  
articles\cite{he90,gel90,gelmk}, and will not be repeated here.

We have carried out the dimer expansions for the $T=0$ ground state energy
per site $E_0/N$,  the antiferromagnetic susceptibility $\chi$,
and the lowest lying triplet excitation spectrum $\Delta (k_x, k_y)$
(odd parity under interchange of the planes)
up to order $(1/y)^{11}$.
The resulting power series in $1/y$ for the ground state energy
per site $E_0/N$ and the antiferromagnetic susceptibility $\chi$ 
are presented in Table I.
Table of series coefficients for the triplet excitation 
spectrum $\Delta (k_x, k_y)$ would require inordinate amount
of space to reproduce in print, it is available from the authors
on request. Instead, we also present in Table I the series 
for the minimum energy gap $m=\Delta (\pi, \pi)$.
The dimer expansion was firstly carried out 
by Hida\cite{hid} up to order  $(1/y)^{6}$
in 1992, and  extended to order $(1/y)^{8}$ by Gelfand\cite{gel}
recently. Our results agree with these previous results, 
and extend the series by three terms. 
A list of 19093 linked clusters of up to 12 sites contribute to
the triplet excitation spectrum.

To determine the critical point $y_c$, we construct the Dlog Pad\'{e}
approximants\cite{gut} to the $\chi$ and $m$ series, 
and since the transition should lie in
the universality class of the classical $d=3$ Heisenberg model
(our analysis also supports this), we expect that the critical 
index for $\chi$ and $m$
should be approximately  1.40 and 0.71, respectively. The 
exponent-biased Dlog Pad\'{e} approximants\cite{gel}
gives the critical point $1/y_c=0.3942(8)$, or $y_c=2.537(5)$.

The spectra for some particular values of $J_1/J_2$ are 
illustrated in Fig. 1,
where the direct sum to the series at $y=y_c$ is indeed consistent
with the Pad\'{e} approximants that one can construct. 
To compute the critical spin-wave velocity $v$, one expands the spectrum
$\Delta$ in the vicinity of wave-vector $(\pi,\pi)$ up to ${\bf k}^2$:
\begin{equation}
\Delta (\pi-k_x, \pi-k_y)[1/y] = C(1/y) + D(1/y) (k_x^2 + k_y^2) + \cdots
\label{eqv}
\end{equation}
and it is easy to prove\cite{gel} that the critical spin-wave velocity is 
equal to $(2CD)^{1/2}$ at $y_c$.
The series coefficients for $2CD$ in $J_1/J_2$ are listed in Table I.  
Using  Pad\'{e} approximants, one can estimate the properties at 
critical point $y_c$: 
the spin-wave velocity is estimated to be
$(v/J_2)^2=0.559(2)$, or $v/J_2=0.748(2)$, 
the ground-state energy per site is $E_0/N=-0.4456(4) J_2$, 
the optical spin-wave gap (the gap at $\Gamma =(0,0)$) 
is $\Delta_{\rm opt} =1.737(1) J_2$,
and the gap at ${\rm X}=(\pi,0)$ is $\Delta_{\rm X} =1.368(8) J_2$,
where the uncertainly is largely associated with the
uncertainly in $y_c$.

\section{Ising expansions}
To construct an expansion about the Ising limit for this system, 
one has to introduce an anisotropy parameter $x$, 
and writes the Hamiltonian for Heisenberg-Ising model as:
\begin{equation}
H/J_1 = H_0 + x V  \label{Hising}
\end{equation}
where 
\begin{eqnarray}
H_0 &= & \sum_{\alpha=1,2} \sum_{\langle i,j\rangle} 
S_{\alpha,i}^z S_{\alpha,j}^z
+ y \sum_{i} S_{1,i}^z S_{2,i}^z + 
 t \sum_{\alpha,i} \epsilon_{\alpha,i} S_{\alpha,i}^z \nonumber \\
V &= & \sum_{\alpha=1,2} \sum_{\langle i,j\rangle} 
 ( S_{\alpha,i}^x S_{\alpha,j}^x + S_{\alpha,i}^y S_{\alpha,j}^y )  + 
y \sum_{i} ( S_{1,i}^x S_{2,i}^x + S_{1,i}^y S_{2,i}^y ) 
- t \sum_{\alpha,i} \epsilon_{\alpha,i} S_{\alpha,i}^z 
\end{eqnarray}
and $\epsilon_{\alpha,i}=\pm 1 $ on the two sublattices. The last term
in both $H_0$ and $V$ is a local staggered field term, which can be
included to improve convergence.
The limits $x=0$ and $x=1$ correspond to the Ising model and
the isotropic Heisenberg model respectively.
The operator $H_0$ is taken as the unperturbed
Hamiltonian, with the unperturbed ground state being the 
usual N\'{e}el state.
The operator $V$ is treated as  a perturbation.
It flips a pair of spins on neighbouring sites. 

The Ising series have been calculated for 
the ground state energy per site $E_0/N$,
the staggered magnetization $M$,  the uniform 
perpendicular susceptibility $\chi_{\perp}$,
and the lowest lying triplet excitation spectrum $\Delta (k_x, k_y, k_z)$
for several ratio of couplings and (simultaneously) for several values of $t$
 up to order $x^{10}$  
(the series for uniform perpendicular susceptibility $\chi_{\perp}$
 is one order less).
The resulting series for $y=0.5, 1, 1.5, 2.5$ and $t=0$  are listed in
Tables II and III (the series coefficients for triplet 
excitation spectrum are listed
only for case of $y=1$), the series for other value of $y$ are 
available on request. 

We can also compute the spin-wave velocity $v$ through a similar expansion 
of $\Delta (k_x,k_y,0)$ in the vicinity of $k_x=k_y=0$
as in Eq.(\ref{eqv}), and the
series coefficients for $2CD$ in $x$ are also listed in Table II.
With this series and the series for the perpendicular susceptibility
$\chi_{\perp}$, one can get the series for the spin-wave 
stiffness $\rho_s$ through
the following ``hydrodynamic'' relation:
\begin{equation}
\rho_s = v^2 \chi_{\perp}~.
\end{equation}

There are two branches of the spin-wave dispersion.
From the series 
one can see
that the symmetric (optical) excitation spectrum  $\Delta (k_x, k_y, 0)$  
(that is, with even  parity under 
interchange of the planes) is related to the antisymmetric
(acoustic) excitation spectrum $\Delta (k_x, k_y, \pi)$ 
(that is, with odd  parity under interchange of the planes) by
\begin{equation}
\Delta (k_x, k_y, 0)= \Delta (\pi-k_y, \pi-k_x, \pi)
\end{equation}
so we only consider the antisymmetric excitation spectrum here.

The conventional linear spin-wave theory\cite{rez,hay,sw}
 predicts the magnon dispersions as:
\begin{equation}
\Delta(k_x,k_y,\pi) =4SJ_1 [ (1-\gamma_k^2 )+y(1+ \gamma_k )/2]^{1/2}
\end{equation}
where $\gamma_k={1\over 2} [\cos (k_x) + \cos (k_y)]$. Therefore 
one can see that the
acoustic spin-wave gap $\Delta_{\rm aco}=\Delta(\pi,\pi,\pi)$ is zero 
even for nonzero $J_2$, the optical gap is
\begin{equation}
\Delta_{\rm opt} = \Delta(0,0,\pi)=4S \sqrt{J_1 J_2}\label{eqopt}
\end{equation}
which is non-zero as long as $J_2\neq 0$, and 
the gap at $X=(\pi,0,\pi)$ is
\begin{equation}
\Delta_{X}=4S J_1 (1+y/2)^{1/2}.
\end{equation}

In a recent paper, Millis and Monien\cite{mil96} got
the following relation for  
the optical spin-wave gap by using the Schwinger boson method:
\begin{equation}
J_2 ={\Delta^2_{\rm opt} \over 16\rho_s +2\Delta_{\rm opt}/\pi} \label{gap_millis}
\end{equation}
and so at the limit of $J_2\to 0$, one results that 
$\Delta_{\rm opt} = 4S\sqrt{J_1 J_2 Z_{\rho}}$, which differs
from Eq.(\ref{eqopt}) by a factor of $\sqrt{Z_{\rho}}$, where
$Z_\rho=\rho_s/(J_1 S^2)\simeq 0.72$.

Having obtained the series we first attempt to identify critical points
and to determine the nature of the phase diagram in the $(x-y)$
plane. Naive Dlog Pad\'{e} analysis reveals lines of singularities,
as shown in Fig. 2. These are obtained consistently for 
the staggered magnetization $M$
and the uniform perpendicular susceptibility $\chi$. We note, 
in particular,
that for the critical point $x_c$ is about equal to 1 at $J_2/J_1=2.55$,
this is consistent with the critical 
point determined by the dimer expansions.

At the next stage of the analysis, we try to extrapolate
 the series to the isotropic point ($x=1$)
for those values of the exchange coupling
parameters which lie within the N\'{e}el-ordered phase at $x=1$.
For this purpose, we first transform the series to a new variable 
\begin{equation}
\delta = 1- (1-x)^{1/2}
\end{equation}
to remove the singularity at $x=1$ predicted by the spin-wave theory. This 
was first proposed by Huse\cite{hus} and was also
used in our earlier work on the square lattice case\cite{zhe1}.
We then use both the integrated first-order inhomogeneous
differential approximants\cite{gut} and the naive Pad\'{e} approximants to
extrapolate the series to the isotropic point $\delta=1$  ($x=1$).
The results for the ground state energy per site $E_0/N$,
the staggered magnetization $M$,  
the uniform perpendicular susceptibility $\chi_{\perp}$,
and spin-wave velocity $v$, and the spin-wave stiffness $\rho_s$
are shown in Figs. 3-5. 
The results for $J_2=0$ presented in the figures 
are taken from our earlier work on the single-layer 
square lattice\cite{zhe1,zhe2}.
The estimates of the ground state energy per site $E_0/N$
from dimer expansions is also shown in Fig. 3.
Note that in the classical spin-wave (large $S$) level, 
the uniform perpendicular 
susceptibility $\chi_{\perp}$, the spin-wave velocity $v$, and 
spin-wave stiffness $\rho_s$ are found to be:
\begin{eqnarray}
\chi_{\perp}^{\rm classical} &=& {1\over 8 J_1} (1-y/4) \nonumber \\
v^{\rm classical} &=& 2 \sqrt{2} S J_1 (1+y/4)^{1/2} \\
\rho_s^{\rm classical} &=& J_1 S^2 (1-y^2/16) \nonumber 
\end{eqnarray}
so in order to remove the trivial dependent of these quantities on
$y$,  Figs. 4-5 have been presented in terms of quantum renormalizations
of the classical spin-wave values.
We note that $M$, $\chi_{\perp}$, and $\rho_s$  first increase 
for small $J_2/J_1$,
pass through a maximum at about $y\simeq 0.5$, and 
then decrease for larger value of $J_2/J_1$,
 and vanishes at about $J_2/J_1=2.55$, which is consistent
with the critical point determined by the dimer expansion.
The reason that in the case of small $J_2/J_1$,
the interlayer coupling enhances the antiferromagnetic long-range order
is the system acquires a weak three dimensionality and 
quantum fluctuation is suppressed. 

The antisymmetric excitation spectra $\Delta (k_x, k_y, \pi)$  
for some particular values of $y$ are illustrated in
Fig. 6, where we can see that the excitation is gapless 
at $(\pi,\pi,\pi)$ point.
In Fig. 7, we present a plot of energy gaps
at $(0,0,\pi)$ and $(\pi,0,\pi)$ obtained from the linear spin-wave 
expansion, the Schwinger boson method\cite{mil96}, the dimer expansions, and the Ising 
expansions. We can see, from Figs. 6-7, that
the spin-wave gap  evolves 
very smoothly from the ordered side
to the disordered side of the transition, (the gap at $X=(0,0,\pi)$ 
turns out to have a small discontinuity at transition point, which may be due to
the estimates from Ising expansions being a little too small).
From Fig. 7, we can also see that for $\Delta_{\rm X}$, the results of 
leading order spin-wave theory is about 10\% smaller than our 
estimates from Ising expansions. 
For optical magnon gap $\Delta_{\rm opt}$,
the Schwinger boson method give consistent results
with those estimated from the Ising expansions over 
the region $0.2 \lesssim y \lesssim 1$, 
but as $y$ decreases from 0.2, the optical magnon gap estimated by the Ising expansions
becomes larger than that predicted by the spin-wave theory, and also by the Schwinger
boson method.

Finally, let us estimate the interlayer exchange constant for
YBa$_2$Cu$_3$O$_{6.2}$ using the optical magnon gap estimated by Ising expansions:
the recent neutron scattering measurements\cite{rez,hay} have report that
the optical magnon gap for YBa$_2$Cu$_3$O$_{6.2}$ is
$\Delta_{\rm opt}\simeq 70$meV, and if
we take the canonical value of $J_1=120$meV, we estimate that $J_2/J_1\simeq
0.07$, whereas $J_2/J_1$ estimated by the spin-wave theory\cite{rez,hay}
and by the Schwinger boson
theory\cite{mil96} are
$0.08$ and $0.14$, respectively.

\section{High temperature series expansions}
We now turn to the thermodynamic properties of the  system
at finite temperatures. We have developed
high-temperature series expansions for the uniform magnetic 
susceptibility $\chi (T)$ 
and the specific heat $C (T)$, for the system with arbitrary $J_1/J_2$,
\begin{eqnarray}
 T \chi (T) &=& {1 \over N} \sum_i \sum_j {{\rm Tr}
 S_i^z S_j^z e^{-\beta H} \over  {\rm Tr} e^{-\beta H } } \nonumber  \\  &&  \\
 C (T) &=& {\partial U \over \partial T} \nonumber
\end{eqnarray}
where $N$ is the number of sites, and $\beta=J_2/(k_BT)$, and the internal
energy $U$ is defined by
\begin{equation}
U =  { {\rm Tr} H e^{-\beta H} \over  {\rm Tr} e^{-\beta H } }
\end{equation}
The series were computed to order $\beta^{9}$ for $\chi(T)$ 
and to order  $\beta^{8}$ for $C(T)$. 
These series are listed in Table IV. We use  the 
integrated first-order inhomogeneous
 differential approximants\cite{gut} and the naive Pad\'{e} approximants
to extrapolate the series. 
The resulting estimates for some particular values of $J_1/J_2$ are
shown in Figs. 8-9. For the susceptibility, in order to remove the trivial dependence of
$\chi(T)$ on Curie-Weiss temperature, we present in Fig. 8 the 
$(1/\chi(T)-T_{\rm cw})/J_1$ as a function of $T/J_1$ where the 
Curie-Weiss temperature $T_{\rm cw}=J_1+J_2/4$.
From this graph, we can see that for $y > y_c$,  the 
susceptibility $\chi (T)$ converge nicely to very low temperature, and
vanishes in the low temperature limit, which consistent with a gapped system.
For $y \lesssim y_c$, the extrapolation, as judged by the
consistency of different approximants, do not converge well in low-temperature
region.
At the critical ratio $(J_2/J_1)_c$, Monte Carlo
simulations\cite{san94} have shown that the universal, linear behavior of
$\chi(T)$ extends up to very high temperature $T\sim J_1$, 
unfortunately our series extrapolation
is not reliable enough to see this behavior.

\section{CONCLUSIONS}
In the preceding sections we have presented and analyzed a 
variety of  high-order perturbation series expansions
for the two-layer, $S={1\over 2}$, square lattice Heisenberg 
(and Heisenberg-Ising) model.
At $T=0$, we have added three more terms to existing series 
for the ground state energy $E_0/N$, 
the antiferromagnetic susceptibility $\chi$,
and the lowest lying triplet excitation spectrum $\Delta (k_x, k_y)$
from the expansions about
the interlayer dimer limits, and also carried out new perturbation 
series expansions  about the Ising limit up to
tenth order for a number of properties.
In addition, high-temperature series for the system with arbitrary $J_1/J_2$
are computed to  order $\beta^{9}$ for uniform magnetic 
susceptibility $\chi(T)$ and to order $\beta^{8}$ for the specific heat $C(T)$.

Based on exponent-biased analyses of the high-order dimer expansions
for triplet energy gap and antiferromagnetic susceptibility, we estimate
the antiferromagnetic order-disorder transition occurring at $T=0$ at a
critical coupling ratio $(J_2/J_1)_c = 2.537(5)$,
which is in good agreement
with the quantum Monte Carlo results\cite{san94} $(J_2/J_1)_c = 2.51(2)$.
The spin-wave velocity at the critical point is estimated to be 
$v/J_2 =0.748(2)$, (equivalently, $v/J_1 = 1.898(9)$), which is slightly higher than
the QMC results\cite{san94} $v/J_1 = 1.69(2)$.
The spin-triplet excitation spectra for the system on both ordered and disordered phases 
are computed, and they seem to be evolve smoothly from the ordered side
to the disordered side of the transition.
At the limit of $J_2/J_1\to 0$, the optical magnon gap obtained by
the Ising expansions is larger than $2\sqrt{J_1 J_2}$, instead of
less than $2\sqrt{J_1 J_2}$ as predicted by the Schwinger boson method,
and this implies the interlayer exchange constant is $J_2\simeq 0.07 J_1$
for YBa$_2$Cu$_3$O$_{6.2}$.

\acknowledgments
This work forms part of a research project supported by a grant 
from the Australian Research Council. I would like to thank Profs. 
J. Oitmaa and C.J. Hamer
for a number of valuble discussions and suggestions.


\begin{figure}[htb]
\caption{Plot of the antisymmetric spin-triplet excitation spectrum 
$\Delta(k_x,k_y)$ along high-symmetry
cuts through the Brillouin zone for the system with coupling ratios
$J_1/J_2=0.1, 0.2, 0.3, 0.3942$ (shown in
the figure from  the top to 
the bottom at $(\pi,\pi)$  respectively), 
the lines are the estimates by direct sum to the dimer series,
and the points (circles with error bar for the case of $J_1/J_2=0.3942$ only) 
are the estimates of the Pad\'{e} approximants to the dimer series.}
\label{fig:fig1}
\end{figure}

\begin{figure}[htb]
\caption{The phase boundary for the Heisenberg-Ising model without
the staggered field ($t=0$)
obtained from analysis of the Ising series for
staggered magnetization $M$ and 
uniform perpendicular susceptibility $\chi_{\perp}$.}
\label{fig:fig2}
\end{figure}

\begin{figure}[htb]
\caption{The rescaled ground-state energy per site $E_0/N$ as 
function of $J_2/(J_1+J_2)$.
The crosses with dashed line connecting them are the estimates from
Ising expansion, and the solid curve is the estimates derived from the dimer
expansions. The error bars are much smaller than
the symbols.
}
\label{fig:fig3}
\end{figure}

\begin{figure}[htb]
\caption{The staggered magnatization $M$ and the
renormalization factor 
of the uniform perpendicular 
susceptibility $Z_{\chi}=8\chi_{\perp} J_1/(1-y/4)$
 (at $T=0$)
{\it versus} $J_2/J_1$ as estimated by  Ising expansions.
}
\label{fig:fig4}
\end{figure}


\begin{figure}[htb]
\caption{The renormalization factor 
of the spin-wave velocity  $Z_c = v/[J_1(2+y/2)^{1/2}]$
and  the spin-wave stiffness 
$Z_{\rho}=4 \rho_s/[J_1 (1-y^2/16)]$ 
{\it versus} $J_2/J_1$ as estimated by Ising expansions
and  dimer expansions (for $Z_c$ at the critical ratio $(J_2/J_1)_c=2.537$ only).}
\label{fig:fig5}
\end{figure}


\begin{figure}[htb]
\caption{Plot of the antisymmetric spin-triplet excitation spectrum 
$\Delta(k_x, k_y, \pi)$
(derived from the Ising expansions) along high-symmetry
cuts through the Brillouin zone for the system with coupling ratios
$J_2/J_1=0.5, 1, 1.5, 2.5$ (shown in
the figure from  the top to 
the bottom at $(\pi,0)$  respectively)}
\label{fig:fig6}
\end{figure}

\begin{figure}[htb]
\caption{The rescaled energy gap at $(0,0,\pi)$ (the optical magnon gap)
and at $X=(\pi,0,\pi)$ as 
function of $J_2/(J_1+J_2)$. The solid curves at large $J_2/(J_1+J_2)$
 are the extrapolation  derived from
the dimer expansions, the crosses with error bars are the estimates
of Ising expansions, the long dashed lines at small $J_2/(J_1+J_2)$ 
are the results of the linear spin-wave theory, the open circles connected
by a short dashed line are the results of Eq.(\protect\ref{gap_millis}) (where
the $\rho_s$ is taken from the results of Fig. 5),
and the vertical dotted line indicates the position of the transition.
}
\label{fig:fig7}
\end{figure}

\begin{figure}[htb]
\caption{Susceptibility $(1/\chi (T)-T_{\rm cw})/J_1$ as a function of temperature
for the system with coupling ratios
$J_2/J_1=5,10/3,2.537,5/3,1$, 
several different integrated  differential approximants to the 
high temperature series are shown for each $J_2/J_1$.
}
\label{fig:fig8}
\end{figure}
 
\begin{figure}[htb]
\caption{The specific heat $C(T)$ as a function of temperature for 
 the system with coupling ratios
$J_1/J_2= 0.1, 0.3$, 
several different integrated  differential approximants to the 
high temperature series are shown for each $J_1/J_2$
}
\label{fig:fig9}
\end{figure}

\widetext
\begin{table}
\squeezetable
\setdec 0.0000000000000
\caption{Series coefficients for dimer expansions of  the
ground-state energy per site $E_0/(NJ_2)$, the
antiferromagnet susceptibility $\chi$, the gap $m/J_2$,
and the critical spin-wave velocity (as described in the text).
Coefficients of $(1/y)^n$
up to order $n=11$ are listed.} \label{tab1}
\begin{tabular}{rrrrr}
 \multicolumn{1}{c}{n} &\multicolumn{1}{c}{$E_0/(NJ_2)$}
&\multicolumn{1}{c}{$J_2 \chi$} &\multicolumn{1}{c}{$m/J_2$} &\multicolumn{1}{c}{$2CD/J_2^2$}  \\
\tableline
  0 &\dec $-$3.750000000$\times 10^{-1}$ &\dec   1.000000000 &\dec   1.000000000 &\dec   0.000000000 \\
  1 &\dec   0.000000000 &\dec   4.000000000 &\dec $-$2.000000000 &\dec   1.000000000 \\
  2 &\dec $-$3.750000000$\times 10^{-1}$ &\dec   1.300000000$\times 10^{1}$ &\dec   0.000000000 &\dec   0.000000000 \\
  3 &\dec $-$1.875000000$\times 10^{-1}$ &\dec   3.950000000$\times 10^{1}$ &\dec $-$1.500000000 &\dec   2.375000000 \\
  4 &\dec $-$1.171875000$\times 10^{-1}$ &\dec   1.128125000$\times 10^{2}$ &\dec $-$1.500000000 &\dec   1.437500000 \\
  5 &\dec   1.289062500$\times 10^{-1}$ &\dec   3.131076389$\times 10^{2}$ &\dec $-$5.468750000$\times 10^{-1}$ &\dec $-$1.320312500 \\
  6 &\dec   2.373046875$\times 10^{-1}$ &\dec   8.535167824$\times 10^{2}$ &\dec   1.191406250$\times 10^{-1}$ &\dec $-$7.851562500$\times 10^{-1}$ \\
  7 &\dec   2.162475586$\times 10^{-1}$ &\dec   2.304592044$\times 10^{3}$ &\dec $-$4.202636719 &\dec $-$2.716064453 \\
  8 &\dec $-$4.346008301$\times 10^{-1}$ &\dec   6.169223926$\times 10^{3}$ &\dec $-$1.060513306$\times 10^{1}$ &\dec   3.307617188 \\
  9 &\dec $-$9.492845535$\times 10^{-1}$ &\dec   1.640000304$\times 10^{4}$ &\dec $-$2.693954849$\times 10^{1}$ &\dec   1.409708977$\times 10^{1}$ \\
 10 &\dec $-$1.231647968 &\dec   4.332290008$\times 10^{4}$ &\dec $-$3.315443579$\times 10^{1}$ &\dec   3.259252389 \\
 11 &\dec   6.411397507$\times 10^{-1}$ &\dec   1.139206022$\times 10^{5}$ &\dec $-$8.044747781$\times 10^{1}$ &\dec   2.037096158$\times 10^{1}$ \\
\end{tabular}
\end{table}
 
\widetext
\begin{table}
\squeezetable
\setdec 0.0000000000000
\caption{Series coefficients  for Ising expansions of the
ground-state energy per site $E_0/(NJ_1)$, the staggered magnetization $M$,
uniform perpendicular susceptibility $\chi_{\perp}$, the energy gap 
$m=\Delta (0,0,0) =\Delta(\pi,\pi,\pi)$,
and the critical spin-wave velocity $2CD$ (as described in the text) for $J_2/J_1=0.5,1,1.5,2.5$ and $t=0$,
Nonzero coefficients $x^n$
up to order $n=10$ are listed.}
 \label{tab2}
\begin{tabular}{rrrrr}
 \multicolumn{1}{c}{n} &\multicolumn{1}{c}{$J_2/J_1=0.5$}
&\multicolumn{1}{c}{$J_2/J_1=1$} &\multicolumn{1}{c}{$J_2/J_1=1.5$} &\multicolumn{1}{c}{$J_2/J_1=2.5$}  \\
\tableline
\multicolumn{5}{c}{$E_0/(J_1N)$} \\
  0 &\dec $-$5.62500000$\times 10^{-1}$ &\dec $-$6.25000000$\times 10^{-1}$ &\dec $-$6.87500000$\times 10^{-1}$ &\dec $-$8.12500000$\times 10^{-1}$ \\
  2 &\dec $-$1.50669643$\times 10^{-1}$ &\dec $-$1.56250000$\times 10^{-1}$ &\dec $-$1.81423611$\times 10^{-1}$ &\dec $-$2.86221591$\times 10^{-1}$ \\
  4 &\dec  1.06782048$\times 10^{-3}$ &\dec  2.32514881$\times 10^{-5}$ &\dec $-$1.82797961$\times 10^{-3}$ &\dec $-$1.01025331$\times 10^{-2}$ \\
  6 &\dec $-$1.55432939$\times 10^{-3}$ &\dec $-$2.51031531$\times 10^{-3}$ &\dec $-$3.56180060$\times 10^{-3}$ &\dec $-$5.93087945$\times 10^{-3}$ \\
  8 &\dec $-$6.14963212$\times 10^{-4}$ &\dec $-$8.46200313$\times 10^{-4}$ &\dec $-$1.01380621$\times 10^{-3}$ &\dec $-$1.82417072$\times 10^{-3}$ \\
 10 &\dec $-$3.42105908$\times 10^{-4}$ &\dec $-$5.02769527$\times 10^{-4}$ &\dec $-$6.24621980$\times 10^{-4}$ &\dec $-$1.03444977$\times 10^{-3}$ \\
\tableline
\multicolumn{5}{c}{$J_1 \chi_{\perp}$}  \\
  0 &\dec  2.22222222$\times 10^{-1}$ &\dec  2.00000000$\times 10^{-1}$ &\dec  1.81818182$\times 10^{-1}$ &\dec  1.53846154$\times 10^{-1}$ \\
  1 &\dec $-$2.81746032$\times 10^{-1}$ &\dec $-$2.50000000$\times 10^{-1}$ &\dec $-$2.29797980$\times 10^{-1}$ &\dec $-$2.08041958$\times 10^{-1}$ \\
  2 &\dec  2.92713075$\times 10^{-1}$ &\dec  2.59090909$\times 10^{-1}$ &\dec  2.37859757$\times 10^{-1}$ &\dec  2.10928368$\times 10^{-1}$ \\
  3 &\dec $-$3.04452465$\times 10^{-1}$ &\dec $-$2.68827110$\times 10^{-1}$ &\dec $-$2.49554660$\times 10^{-1}$ &\dec $-$2.34325217$\times 10^{-1}$ \\
  4 &\dec  3.05483508$\times 10^{-1}$ &\dec  2.69863988$\times 10^{-1}$ &\dec  2.49828011$\times 10^{-1}$ &\dec  2.29571097$\times 10^{-1}$ \\
  5 &\dec $-$3.09716015$\times 10^{-1}$ &\dec $-$2.74883091$\times 10^{-1}$ &\dec $-$2.56566684$\times 10^{-1}$ &\dec $-$2.43161906$\times 10^{-1}$ \\
  6 &\dec  3.10814223$\times 10^{-1}$ &\dec  2.76075768$\times 10^{-1}$ &\dec  2.57806919$\times 10^{-1}$ &\dec  2.44310355$\times 10^{-1}$ \\
  7 &\dec $-$3.13649358$\times 10^{-1}$ &\dec $-$2.79129287$\times 10^{-1}$ &\dec $-$2.61531350$\times 10^{-1}$ &\dec $-$2.53169737$\times 10^{-1}$ \\
  8 &\dec  3.14366380$\times 10^{-1}$ &\dec  2.79764322$\times 10^{-1}$ &\dec  2.62062184$\times 10^{-1}$ &\dec  2.52501211$\times 10^{-1}$ \\
  9 &\dec $-$3.16219779$\times 10^{-1}$ &\dec $-$2.81901242$\times 10^{-1}$ &\dec $-$2.64657763$\times 10^{-1}$ &\dec $-$2.58530073$\times 10^{-1}$ \\
\tableline
\multicolumn{5}{c}{$M$}  \\
  0 &\dec  5.00000000$\times 10^{-1}$ &\dec  5.00000000$\times 10^{-1}$ &\dec  5.00000000$\times 10^{-1}$ &\dec  5.00000000$\times 10^{-1}$ \\
  2 &\dec $-$8.55389031$\times 10^{-2}$ &\dec $-$7.81250000$\times 10^{-2}$ &\dec $-$8.45389660$\times 10^{-2}$ &\dec $-$1.30714101$\times 10^{-1}$ \\
  4 &\dec $-$8.46292613$\times 10^{-3}$ &\dec $-$8.53938581$\times 10^{-3}$ &\dec $-$1.34925210$\times 10^{-2}$ &\dec $-$4.35583219$\times 10^{-2}$ \\
  6 &\dec $-$5.85842608$\times 10^{-3}$ &\dec $-$8.07527219$\times 10^{-3}$ &\dec $-$1.15477220$\times 10^{-2}$ &\dec $-$2.85436734$\times 10^{-2}$ \\
  8 &\dec $-$3.93110760$\times 10^{-3}$ &\dec $-$4.86105744$\times 10^{-3}$ &\dec $-$6.30805266$\times 10^{-3}$ &\dec $-$1.93128872$\times 10^{-2}$ \\
 10 &\dec $-$2.73750635$\times 10^{-3}$ &\dec $-$3.51123409$\times 10^{-3}$ &\dec $-$4.48806496$\times 10^{-3}$ &\dec $-$1.37095158$\times 10^{-2}$ \\
\tableline
\multicolumn{5}{c}{$m/J_1$}  \\
  0 &\dec  2.25000000   &\dec  2.50000000   &\dec  2.75000000   &\dec  3.25000000   \\
  2 &\dec $-$1.64330357   &\dec $-$1.77083333   &\dec $-$1.98338294   &\dec $-$2.55445076   \\
  4 &\dec  1.20240301$\times 10^{-1}$ &\dec  5.95393105$\times 10^{-2}$ &\dec  9.76911256$\times 10^{-2}$ &\dec  3.91257865$\times 10^{-1}$ \\
  6 &\dec $-$2.91150072$\times 10^{-1}$ &\dec $-$2.85968590$\times 10^{-1}$ &\dec $-$3.69547989$\times 10^{-1}$ &\dec $-$9.18026495$\times 10^{-1}$ \\
  8 &\dec  8.98232628$\times 10^{-2}$ &\dec  6.28691338$\times 10^{-2}$ &\dec  1.56753567$\times 10^{-1}$ &\dec  1.10864903   \\
 10 &\dec $-$1.90221034$\times 10^{-1}$ &\dec $-$1.73683175$\times 10^{-1}$ &\dec $-$3.02886458$\times 10^{-1}$ &\dec $-$2.00750062   \\
\tableline
\multicolumn{5}{c}{$2CD/J_1^2$}  \\
  2 &\dec  3.97500000   &\dec  4.16666667   &\dec  4.51785714   &\dec  5.59722222   \\
  4 &\dec $-$1.62276149   &\dec $-$1.51750579   &\dec $-$1.78959346   &\dec $-$3.20894532   \\
  6 &\dec  1.30765116   &\dec  1.08647250   &\dec  1.39046692   &\dec  3.71475815   \\
  8 &\dec $-$1.40693583   &\dec $-$1.13102684   &\dec $-$1.70310223   &\dec $-$7.26139043   \\
 10 &\dec  1.60579536   &\dec  1.23771527   &\dec  2.19444652   &\dec  1.47488833$\times 10^{1}$ \\
\end{tabular}
\end{table}

\setdec 0.0000000000
\begin{table}
\squeezetable
\caption{Series coefficients for the Ising expansions of the triplet
spin-wave excitation spectrum $\Delta (k_x, k_y, k_z) =$ 
$ J_1 \sum_{k,n,m,p} a_{k,n,m,p} x^{k} \cos(p k_z)
 [\cos (m k_x) \cos (n k_y) + \cos (n k_x) \cos (m k_y) ]/2$.
 Nonzero coefficients $a_{k,n,m,p}$
up to order $k=10$ for the case of $J_2=J_1$ are listed.}\label{tabisigap}
\begin{tabular}{rr|rr|rr|rr}
\multicolumn{1}{c}{(k,n,m,p)} &\multicolumn{1}{c|}{$a_{k,n,m,p}$}
&\multicolumn{1}{c}{(k,n,m,p)} &\multicolumn{1}{c|}{$a_{k,n,m,p}$}
&\multicolumn{1}{c}{(k,n,m,p)} &\multicolumn{1}{c|}{$a_{k,n,m,p}$}
&\multicolumn{1}{c}{(k,n,m,p)} &\multicolumn{1}{c}{$a_{k,n,m,p}$} \\
\hline
 ( 0,0,0,0) &\dec  2.50000000   &( 6,1,1,0) &\dec $-$5.01421733$\times 10^{-2}$ &( 6,3,3,0) &\dec $-$3.50360695$\times 10^{-3}$ &(10,7,0,1) &\dec $-$1.95483982$\times 10^{-4}$ \\
 ( 2,0,0,0) &\dec $-$1.04166667$\times 10^{-1}$ &( 8,1,1,0) &\dec  3.13641370$\times 10^{-2}$ &( 8,3,3,0) &\dec $-$2.76765928$\times 10^{-3}$ &(10,9,0,1) &\dec $-$3.18431828$\times 10^{-6}$ \\
 ( 4,0,0,0) &\dec  8.82430142$\times 10^{-2}$ &(10,1,1,0) &\dec $-$2.49468947$\times 10^{-2}$ &(10,3,3,0) &\dec $-$2.34255921$\times 10^{-3}$ &( 4,2,1,1) &\dec $-$1.18055556$\times 10^{-1}$ \\
 ( 6,0,0,0) &\dec $-$2.63720521$\times 10^{-2}$ &( 4,3,1,0) &\dec $-$3.93518519$\times 10^{-2}$ &( 8,5,3,0) &\dec $-$7.65220675$\times 10^{-4}$ &( 6,2,1,1) &\dec $-$2.64155861$\times 10^{-2}$ \\
 ( 8,0,0,0) &\dec  1.81429057$\times 10^{-2}$ &( 6,3,1,0) &\dec $-$1.87035451$\times 10^{-2}$ &(10,5,3,0) &\dec $-$1.48448940$\times 10^{-3}$ &( 8,2,1,1) &\dec  5.61540297$\times 10^{-3}$ \\
 (10,0,0,0) &\dec $-$9.88646062$\times 10^{-3}$ &( 8,3,1,0) &\dec $-$3.19610735$\times 10^{-3}$ &(10,7,3,0) &\dec $-$7.64236388$\times 10^{-5}$ &(10,2,1,1) &\dec $-$2.48059104$\times 10^{-2}$ \\
 ( 2,2,0,0) &\dec $-$3.33333333$\times 10^{-1}$ &(10,3,1,0) &\dec $-$1.18183470$\times 10^{-2}$ &( 8,4,4,0) &\dec $-$4.78262922$\times 10^{-4}$ &( 6,4,1,1) &\dec $-$1.05108208$\times 10^{-2}$ \\
 ( 4,2,0,0) &\dec  1.52199074$\times 10^{-2}$ &( 6,5,1,0) &\dec $-$2.10216417$\times 10^{-3}$ &(10,4,4,0) &\dec $-$8.91369509$\times 10^{-4}$ &( 8,4,1,1) &\dec $-$5.39907287$\times 10^{-3}$ \\
 ( 6,2,0,0) &\dec $-$3.34161849$\times 10^{-2}$ &( 8,5,1,0) &\dec $-$2.49774309$\times 10^{-3}$ &(10,6,4,0) &\dec $-$1.33741368$\times 10^{-4}$ &(10,4,1,1) &\dec $-$5.30710289$\times 10^{-3}$ \\
 ( 8,2,0,0) &\dec  1.25625535$\times 10^{-2}$ &(10,5,1,0) &\dec $-$2.84123845$\times 10^{-3}$ &(10,5,5,0) &\dec $-$8.02448208$\times 10^{-5}$ &( 8,6,1,1) &\dec $-$7.65220675$\times 10^{-4}$ \\
 (10,2,0,0) &\dec $-$1.75733942$\times 10^{-2}$ &( 8,7,1,0) &\dec $-$1.09317239$\times 10^{-4}$ &( 2,1,0,1) &\dec $-$6.66666667$\times 10^{-1}$ &(10,6,1,1) &\dec $-$1.16719696$\times 10^{-3}$ \\
 ( 4,4,0,0) &\dec $-$4.91898148$\times 10^{-3}$ &(10,7,1,0) &\dec $-$3.13087047$\times 10^{-4}$ &( 4,1,0,1) &\dec  6.74768519$\times 10^{-2}$ &(10,8,1,1) &\dec $-$5.73177291$\times 10^{-5}$ \\
 ( 6,4,0,0) &\dec $-$4.91156561$\times 10^{-3}$ &(10,9,1,0) &\dec $-$6.36863657$\times 10^{-6}$ &( 6,1,0,1) &\dec $-$5.84006701$\times 10^{-2}$ &( 6,3,2,1) &\dec $-$2.10216417$\times 10^{-2}$ \\
 ( 8,4,0,0) &\dec $-$2.30378946$\times 10^{-3}$ &( 4,2,2,0) &\dec $-$2.95138889$\times 10^{-2}$ &( 8,1,0,1) &\dec  3.34318983$\times 10^{-2}$ &( 8,3,2,1) &\dec $-$5.04717657$\times 10^{-3}$ \\
 (10,4,0,0) &\dec $-$3.17090252$\times 10^{-3}$ &( 6,2,2,0) &\dec $-$1.09825781$\times 10^{-2}$ &(10,1,0,1) &\dec $-$3.12876108$\times 10^{-2}$ &(10,3,2,1) &\dec $-$7.50314475$\times 10^{-3}$ \\
 ( 6,6,0,0) &\dec $-$1.75180347$\times 10^{-4}$ &( 8,2,2,0) &\dec $-$7.99539319$\times 10^{-4}$ &( 4,3,0,1) &\dec $-$1.96759259$\times 10^{-2}$ &( 8,5,2,1) &\dec $-$2.29566203$\times 10^{-3}$ \\
 ( 8,6,0,0) &\dec $-$3.97833036$\times 10^{-4}$ &(10,2,2,0) &\dec $-$7.65794740$\times 10^{-3}$ &( 6,3,0,1) &\dec $-$1.30043279$\times 10^{-2}$ &(10,5,2,1) &\dec $-$2.08196519$\times 10^{-3}$ \\
 (10,6,0,0) &\dec $-$6.37350643$\times 10^{-4}$ &( 6,4,2,0) &\dec $-$5.25541042$\times 10^{-3}$ &( 8,3,0,1) &\dec $-$9.25242372$\times 10^{-4}$ &(10,7,2,1) &\dec $-$2.29270916$\times 10^{-4}$ \\
 ( 8,8,0,0) &\dec $-$6.83232746$\times 10^{-6}$ &( 8,4,2,0) &\dec $-$4.57905354$\times 10^{-3}$ &(10,3,0,1) &\dec $-$6.65895455$\times 10^{-3}$ &( 8,4,3,1) &\dec $-$3.82610338$\times 10^{-3}$ \\
 (10,8,0,0) &\dec $-$3.37994352$\times 10^{-5}$ &(10,4,2,0) &\dec $-$4.19352456$\times 10^{-3}$ &( 6,5,0,1) &\dec $-$1.05108208$\times 10^{-3}$ &(10,4,3,1) &\dec $-$2.63082695$\times 10^{-3}$ \\
 (10,10,0,0)&\dec $-$3.18431828$\times 10^{-7}$ &( 8,6,2,0) &\dec $-$3.82610338$\times 10^{-4}$ &( 8,5,0,1) &\dec $-$1.65065868$\times 10^{-3}$ &(10,6,3,1) &\dec $-$5.34965472$\times 10^{-4}$ \\
 ( 2,1,1,0) &\dec $-$6.66666667$\times 10^{-1}$ &(10,6,2,0) &\dec $-$8.44976702$\times 10^{-4}$ &(10,5,0,1) &\dec $-$1.45569445$\times 10^{-3}$ &(10,5,4,1) &\dec $-$8.02448208$\times 10^{-4}$ \\
 ( 4,1,1,0) &\dec  1.00115741$\times 10^{-1}$ &(10,8,2,0) &\dec $-$2.86588646$\times 10^{-5}$ &( 8,7,0,1) &\dec $-$5.46586197$\times 10^{-5}$  \\
\end{tabular}
\end{table}

\setdec 0.00000000000
\begin{table}
\squeezetable
\caption{Series coefficients for the high-temperature expansions of the 
uniform susceptibility $ \chi (T)= T^{-1} \sum_{m,n} a_{m,n} (J_1/J_2)^m \beta^n$ and
the specific heat $C(T)=\beta^2 \sum_{m,n} a_{m,n} (J_1/J_2)^m \beta^n$.
 Nonzero coefficients $a_{m,n}$
up to order $n=9$ for $\chi$ or order $n=8$ for $C$ are listed.}\label{tabht}
\begin{tabular}{rr|rr|rr|rr}
\multicolumn{1}{c}{(m,n)} &\multicolumn{1}{c|}{$a_{m,n}$}
&\multicolumn{1}{c}{(m,n)} &\multicolumn{1}{c|}{$a_{m,n}$}
&\multicolumn{1}{c}{(m,n)} &\multicolumn{1}{c|}{$a_{m,n}$}
&\multicolumn{1}{c}{(m,n)} &\multicolumn{1}{c}{$a_{m,n}$} \\
\hline
\multicolumn{8}{c}{$\chi (T)$} \\
( 0, 0) &\dec  2.500000000$\times 10^{-1}$ &( 4, 4) &\dec  1.692708333$\times 10^{-2}$ &( 0, 7) &\dec $-$3.986661396$\times 10^{-5}$ &( 6, 8) &\dec  1.061827644$\times 10^{-2}$ \\
( 0, 1) &\dec $-$6.250000000$\times 10^{-2}$ &( 0, 5) &\dec  2.115885417$\times 10^{-4}$ &( 1, 7) &\dec  5.113389757$\times 10^{-4}$ &( 7, 8) &\dec  4.403056796$\times 10^{-3}$ \\
( 1, 1) &\dec $-$2.500000000$\times 10^{-1}$ &( 1, 5) &\dec $-$3.580729167$\times 10^{-3}$ &( 2, 7) &\dec $-$2.441406250$\times 10^{-3}$ &( 8, 8) &\dec  3.978426494$\times 10^{-4}$ \\
( 0, 2) &\dec $-$1.562500000$\times 10^{-2}$ &( 2, 5) &\dec  1.822916667$\times 10^{-2}$ &( 3, 7) &\dec  4.155815972$\times 10^{-3}$ &( 0, 9) &\dec  3.899804709$\times 10^{-6}$ \\
( 1, 2) &\dec  1.250000000$\times 10^{-1}$ &( 3, 5) &\dec $-$2.864583333$\times 10^{-2}$ &( 4, 7) &\dec  6.287977431$\times 10^{-3}$ &( 1, 9) &\dec $-$4.881601485$\times 10^{-5}$ \\
( 2, 2) &\dec  1.250000000$\times 10^{-1}$ &( 4, 5) &\dec $-$4.492187500$\times 10^{-2}$ &( 5, 7) &\dec $-$1.420898438$\times 10^{-2}$ &( 2, 9) &\dec  2.027723524$\times 10^{-4}$ \\
( 0, 3) &\dec  1.302083333$\times 10^{-3}$ &( 5, 5) &\dec $-$9.244791667$\times 10^{-3}$ &( 6, 7) &\dec $-$1.158582899$\times 10^{-2}$ &( 3, 9) &\dec $-$1.131895358$\times 10^{-4}$ \\
( 1, 3) &\dec  1.562500000$\times 10^{-2}$ &( 0, 6) &\dec $-$1.044379340$\times 10^{-4}$ &( 7, 7) &\dec  6.285652282$\times 10^{-4}$ &( 4, 9) &\dec $-$1.836698017$\times 10^{-3}$ \\
( 2, 3) &\dec $-$1.250000000$\times 10^{-1}$ &( 1, 6) &\dec  5.533854167$\times 10^{-4}$ &( 0, 8) &\dec  1.513768756$\times 10^{-6}$ &( 5, 9) &\dec  3.601243761$\times 10^{-3}$ \\
( 3, 3) &\dec $-$4.166666667$\times 10^{-2}$ &( 2, 6) &\dec  1.399739583$\times 10^{-3}$ &( 1, 8) &\dec  3.700861855$\times 10^{-5}$ &( 6, 9) &\dec  3.023339327$\times 10^{-4}$ \\
( 0, 4) &\dec  1.627604167$\times 10^{-3}$ &( 3, 6) &\dec $-$1.527777778$\times 10^{-2}$ &( 2, 8) &\dec $-$6.515260727$\times 10^{-4}$ &( 7, 9) &\dec $-$7.574656653$\times 10^{-3}$ \\
( 1, 4) &\dec $-$1.041666667$\times 10^{-2}$ &( 4, 6) &\dec  2.438151042$\times 10^{-2}$ &( 3, 8) &\dec  3.424169147$\times 10^{-3}$ &( 8, 9) &\dec $-$1.439339774$\times 10^{-3}$ \\
( 2, 4) &\dec  7.812500000$\times 10^{-3}$ &( 5, 6) &\dec  2.350260417$\times 10^{-2}$ &( 4, 8) &\dec $-$5.407569522$\times 10^{-3}$ &( 9, 9) &\dec $-$5.211347415$\times 10^{-4}$ \\
( 3, 4) &\dec  8.333333333$\times 10^{-2}$ &( 6, 6) &\dec  1.991102431$\times 10^{-3}$ &( 5, 8) &\dec $-$1.251705109$\times 10^{-3}$     \\
\hline
\multicolumn{8}{c}{$C(T)$} \\
( 0, 0) &\dec  9.375000000$\times 10^{-2}$ &( 2, 4) &\dec $-$1.025390625$\times 10^{-2}$ &( 3, 6) &\dec  3.450520833$\times 10^{-3}$ &( 9, 7) &\dec $-$6.530006045$\times 10^{-2}$ \\
( 2, 0) &\dec  3.750000000$\times 10^{-1}$ &( 3, 4) &\dec $-$1.367187500$\times 10^{-2}$ &( 4, 6) &\dec $-$4.157307943$\times 10^{-2}$ &( 0, 8) &\dec $-$5.264736357$\times 10^{-5}$ \\
( 0, 1) &\dec  4.687500000$\times 10^{-2}$ &( 4, 4) &\dec  1.464843750$\times 10^{-1}$ &( 5, 6) &\dec  2.871093750$\times 10^{-2}$ &( 2, 8) &\dec $-$5.870092483$\times 10^{-4}$ \\
( 3, 1) &\dec  1.875000000$\times 10^{-1}$ &( 6, 4) &\dec  3.173828125$\times 10^{-2}$ &( 6, 6) &\dec $-$8.115234375$\times 10^{-2}$ &( 3, 8) &\dec $-$4.486810593$\times 10^{-4}$ \\
( 0, 2) &\dec $-$5.859375000$\times 10^{-3}$ &( 0, 5) &\dec  9.399414063$\times 10^{-4}$ &( 8, 6) &\dec  1.951700846$\times 10^{-2}$ &( 4, 8) &\dec  3.243491763$\times 10^{-3}$ \\
( 2, 2) &\dec $-$4.687500000$\times 10^{-2}$ &( 2, 5) &\dec  4.785156250$\times 10^{-3}$ &( 0, 7) &\dec $-$1.816522507$\times 10^{-5}$ &( 5, 8) &\dec $-$1.445297968$\times 10^{-2}$ \\
( 4, 2) &\dec $-$1.640625000$\times 10^{-1}$ &( 3, 5) &\dec  3.759765625$\times 10^{-3}$ &( 2, 7) &\dec  7.376534598$\times 10^{-4}$ &( 6, 8) &\dec  2.286320641$\times 10^{-2}$ \\
( 0, 3) &\dec $-$9.765625000$\times 10^{-3}$ &( 4, 5) &\dec  4.990234375$\times 10^{-2}$ &( 3, 7) &\dec  2.947126116$\times 10^{-4}$ &( 7, 8) &\dec $-$3.248014904$\times 10^{-2}$ \\
( 2, 3) &\dec $-$3.906250000$\times 10^{-2}$ &( 5, 5) &\dec  9.365234375$\times 10^{-2}$ &( 4, 7) &\dec $-$2.781633650$\times 10^{-2}$ &( 8, 8) &\dec  3.321502322$\times 10^{-2}$ \\
( 3, 3) &\dec $-$3.906250000$\times 10^{-2}$ &( 7, 5) &\dec  1.268066406$\times 10^{-1}$ &( 5, 7) &\dec $-$2.221854074$\times 10^{-2}$ &(10, 8) &\dec $-$2.914089021$\times 10^{-2}$ \\
( 5, 3) &\dec $-$1.953125000$\times 10^{-1}$ &( 0, 6) &\dec  4.185994466$\times 10^{-4}$ &( 6, 7) &\dec $-$2.977120536$\times 10^{-2}$  \\
( 0, 4) &\dec $-$1.586914062$\times 10^{-3}$ &( 2, 6) &\dec  4.451497396$\times 10^{-3}$ &( 7, 7) &\dec $-$1.089634487$\times 10^{-1}$  \\
\end{tabular}
\end{table}


\begin{references}
\bibitem[*]{byline1}e-mail address: w.zheng@unsw.edu.au
\bibitem{rez}D. Reznik, {\it et. al.}, Phys. Rev. B{\bf 53}, 14741(1996).
\bibitem{hay}S.M. Hayden, G. Aeppli, T.G. Perring,
H.A. Mook and F. Dogan, Phys. Rev. B{\bf 54}, R6905(1996).
\bibitem{mil96}A.J. Millis and H. Monien, Phys. Rev. B{\bf 54}, 16172(1996).
\bibitem{hid}K. Hida, J. Phys. Soc. Jpn. {\bf 61}, 1013(1992).
\bibitem{gel}M. P. Gelfand, Phys. Rev. B {\bf 53}, 11309(1996).
\bibitem{san94}A.W. Sandvik and D.J. Scalapino,
Phys. Rev. Lett. {\bf 72}, 2777(1994).
\bibitem{san95}A.W. Sandvik, A.V. Chubukov, and S. Sachdev,
Phys. Rev. B {\bf 51}, 16483(1995).
\bibitem{san96}A.W. Sandvik and D.J. Scalapino,
Phys. Rev. B {\bf 53}, R526(1996).
\bibitem{mil}A.J. Millis and H. Monien, Phys. Rev. Lett. {\bf 70},
2810(1993); Phys. Rev. B {\bf 50}, 16606(1994).
\bibitem{chu95}A.V. Chubukov and D.K. Morr,
Phys. Rev. B {\bf 52}, 3521(1995).
\bibitem{miy}T. Miyazaki, I. Nakamura, and D. Yoshioka,
Phys. Rev. B {\bf 53}, 12206(1996).
\bibitem{he90}H.X. He, C.J. Hamer and J. Oitmaa, J. Phys. A {\bf 23}, 1775(1990).
\bibitem{gel90}M. P. Gelfand, R.R.P. Singh, and D.A, Huse, J. of Stat. Phys. {\bf 59},
1093(1990).
\bibitem{gelmk}M. P. Gelfand, Solid State Commun. {\bf 98}, 11(1996).
\bibitem{gut}A.J. Guttmann, in {\it Phase Transitions and Critical Phenomena}, 
edited by C. Domb and J.L. Lebowitz (Academic, New York, 1989), Vol. 13.
\bibitem{sw}T. Matsuda and K. Hida, J. Phys. Soc. Jpn. {\bf 59}, 2223(1990);
K. Hida, J. Phys. Soc. Jpn. {\bf 59}, 2230(1990).
\bibitem{hus}D.A. Huse, Phys. Rev. B{\bf 37}, 2380(1988).
\bibitem{zhe1}W.H. Zheng, J. Oitmaa and C.J. Hamer, Phys. Rev. B {\bf 43},
8321(1991).
\bibitem{zhe2}C.J. Hamer, W.H. Zheng, and J. Oitmaa, Phys. Rev.
B {\bf 50}, 6877(1994).
\end{references}
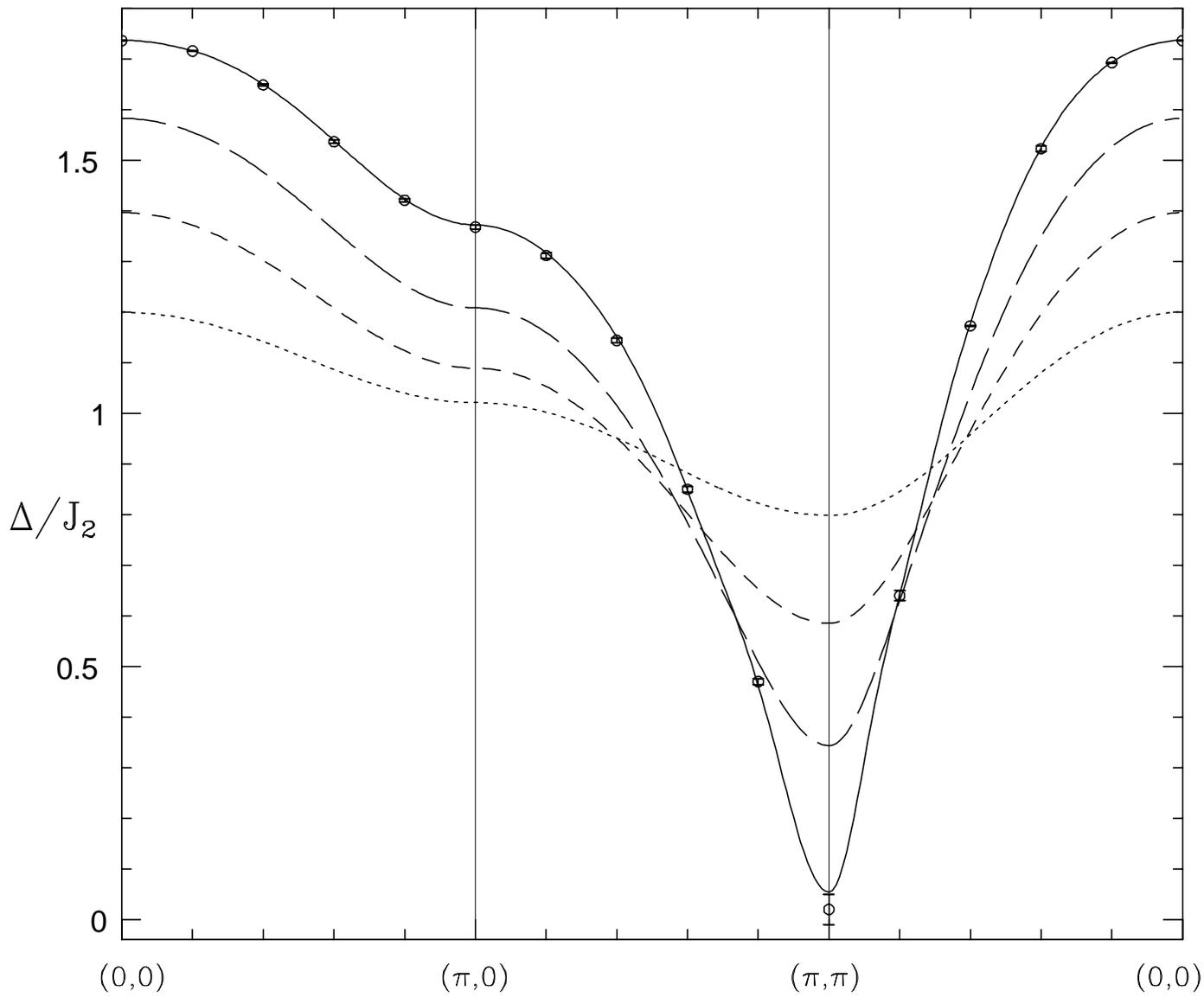
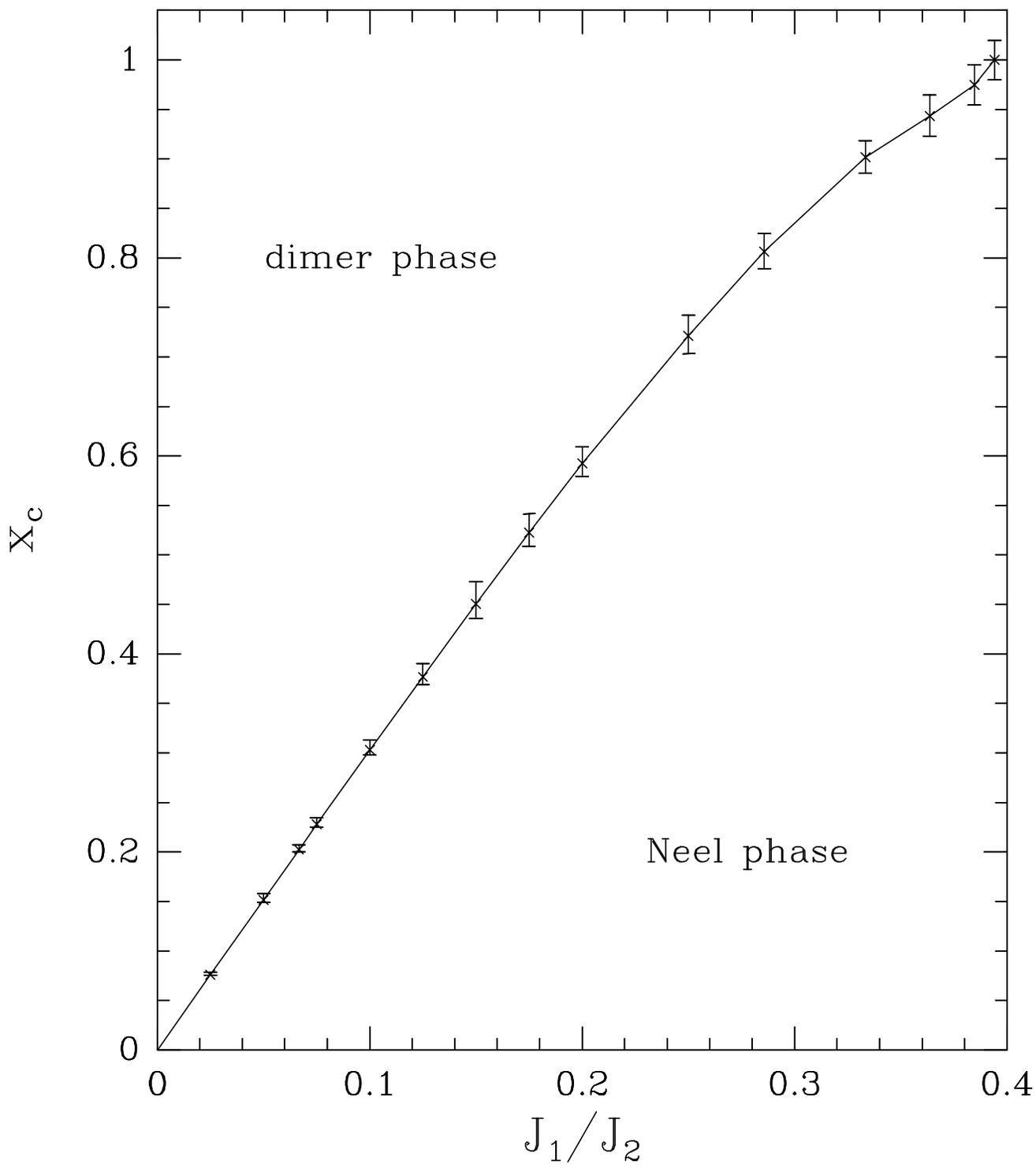
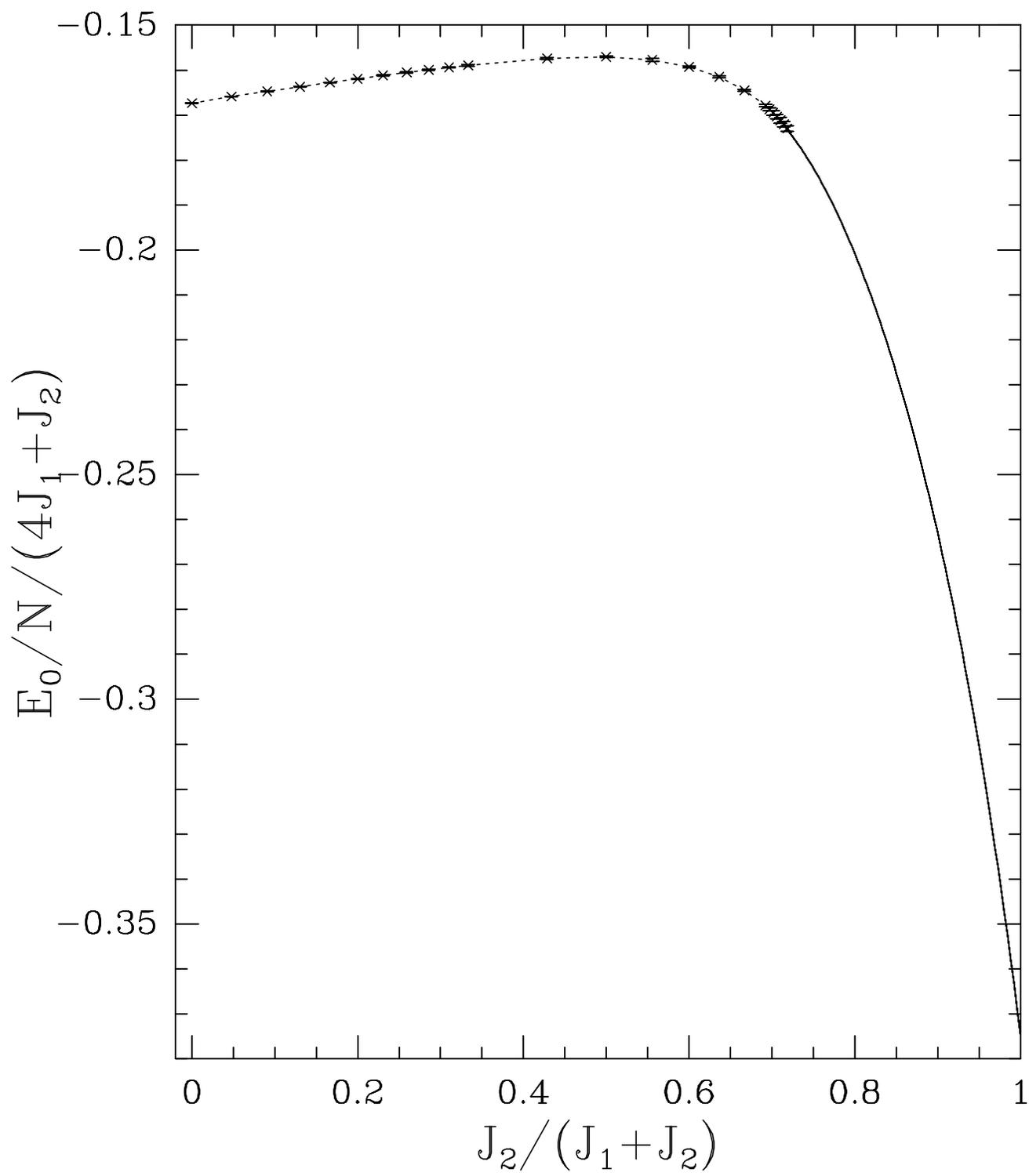
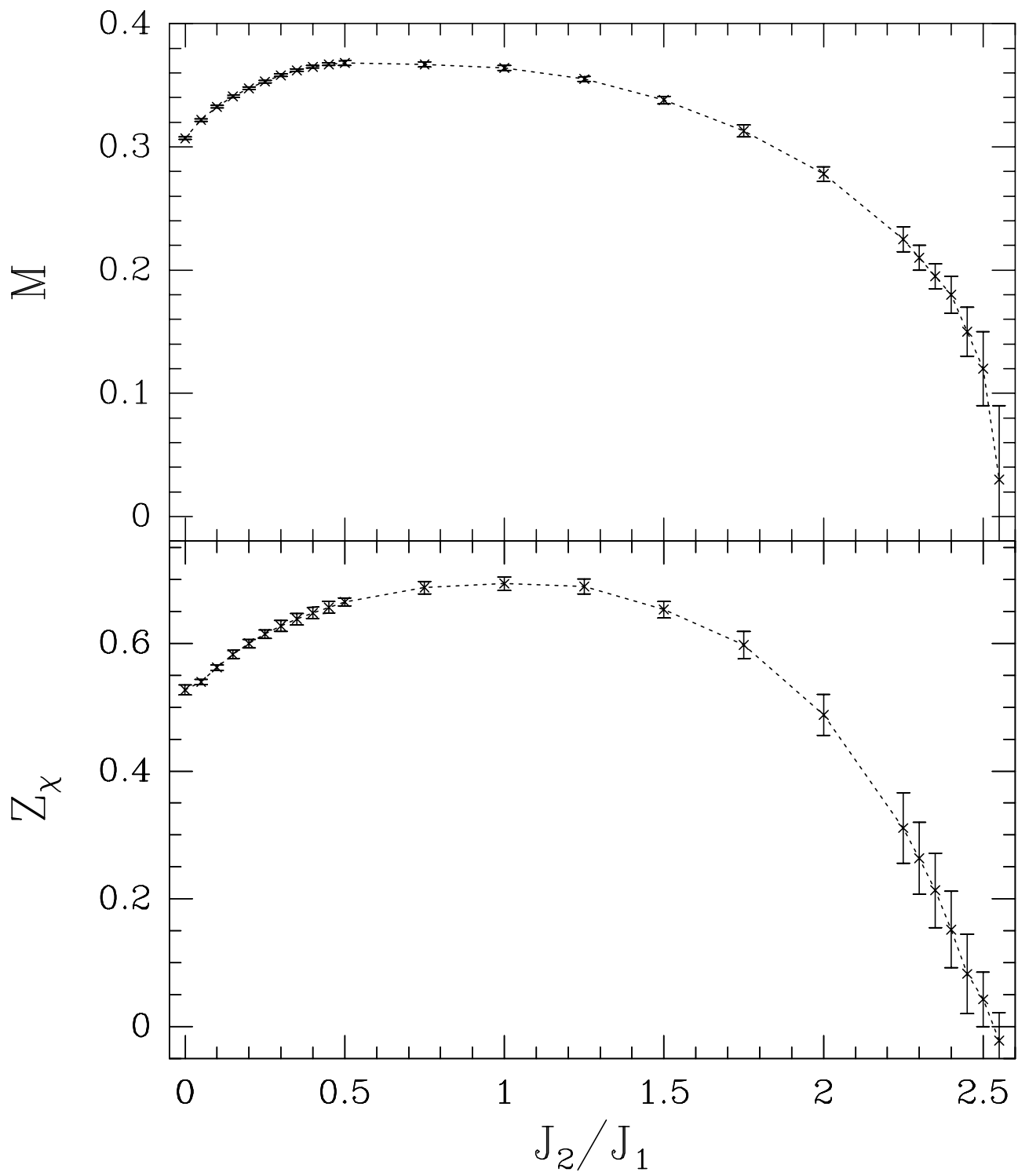
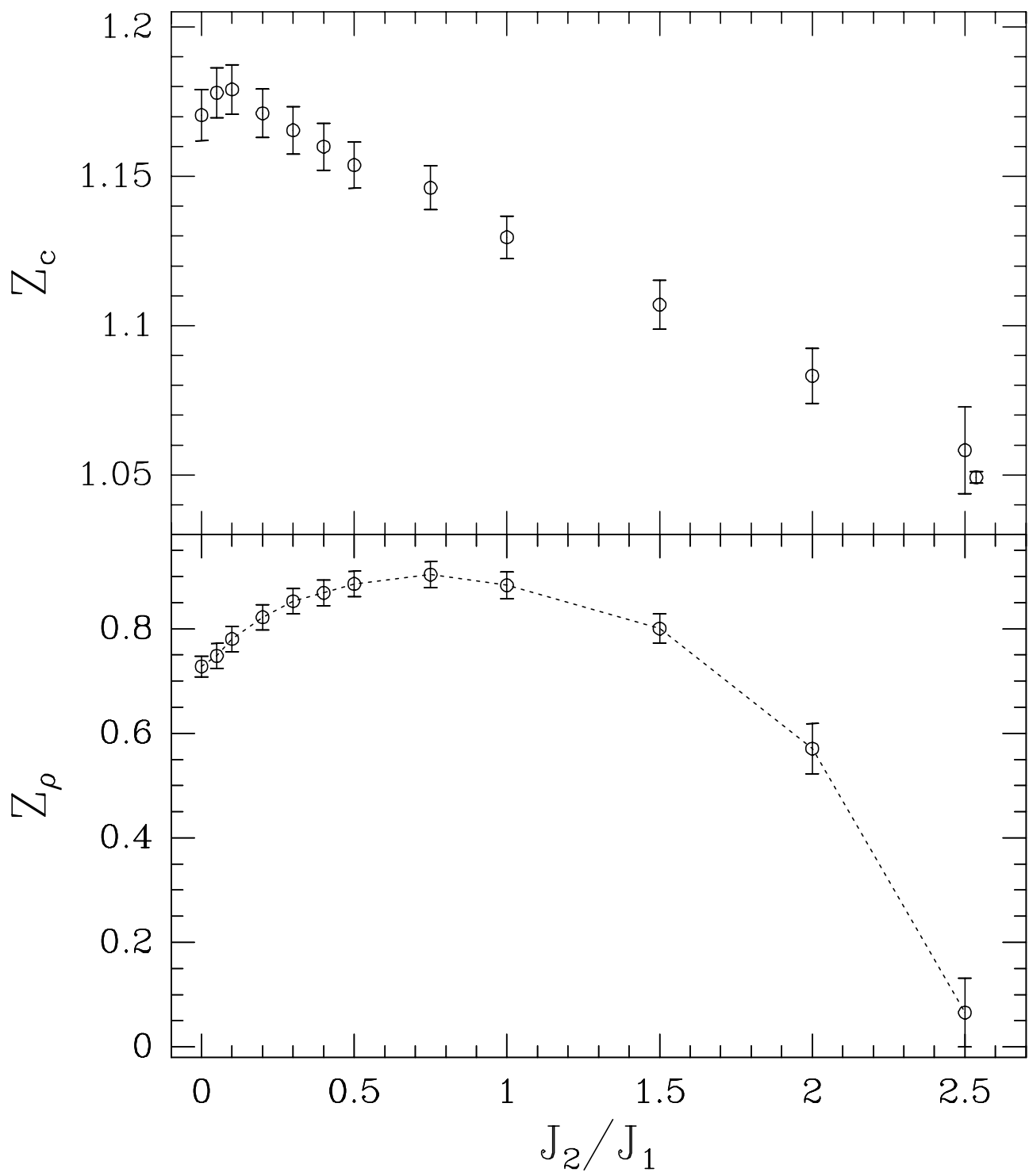
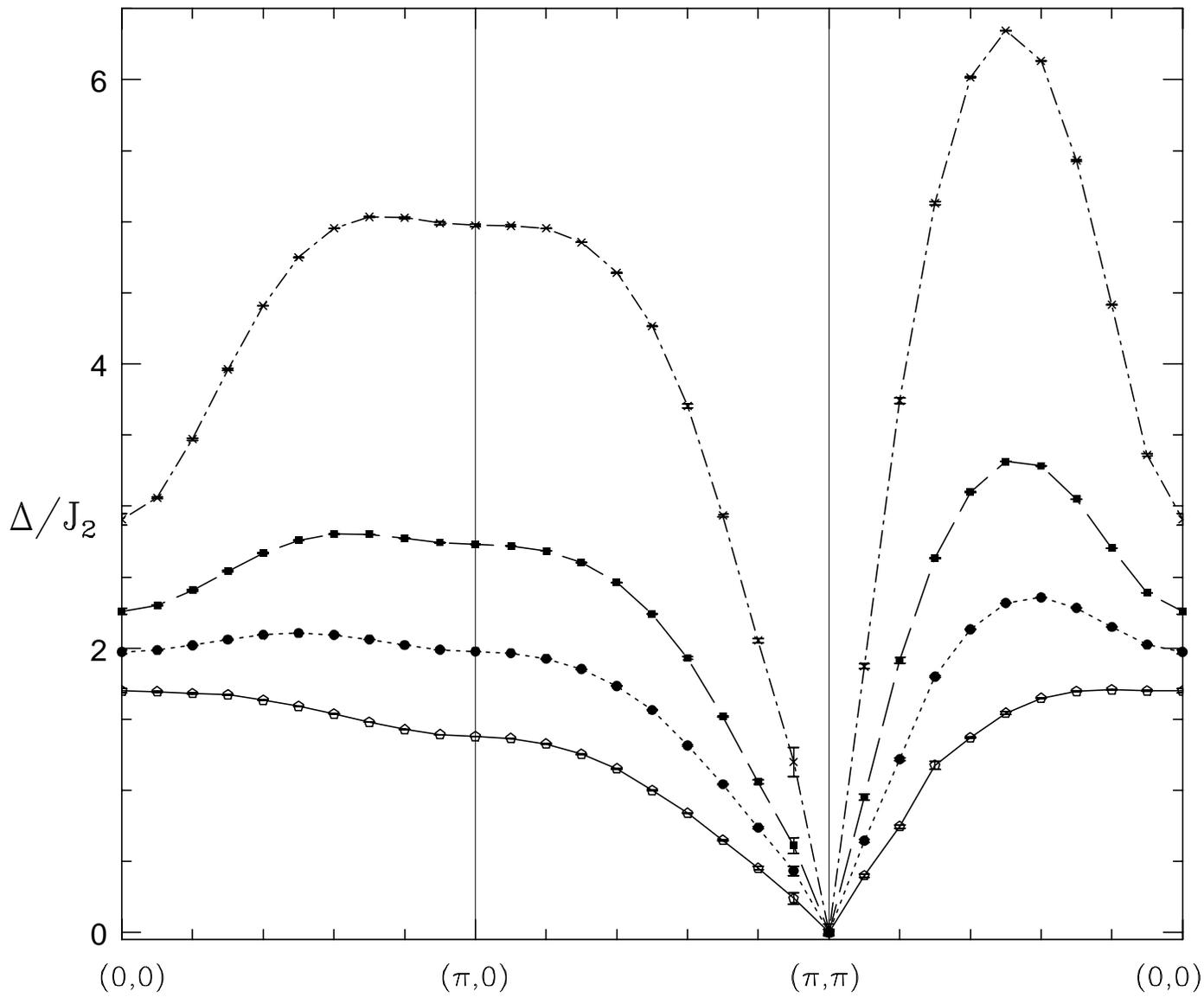
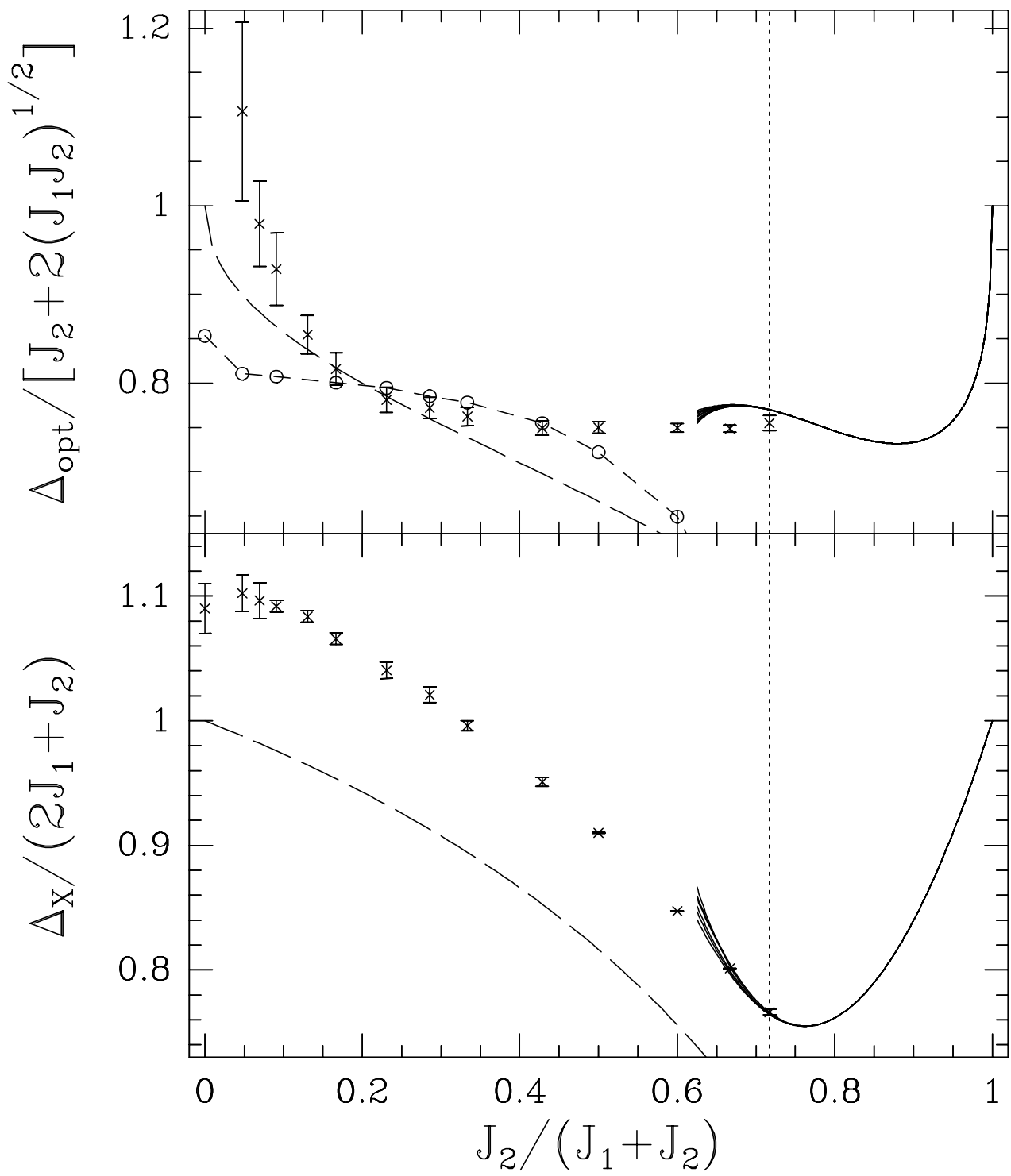
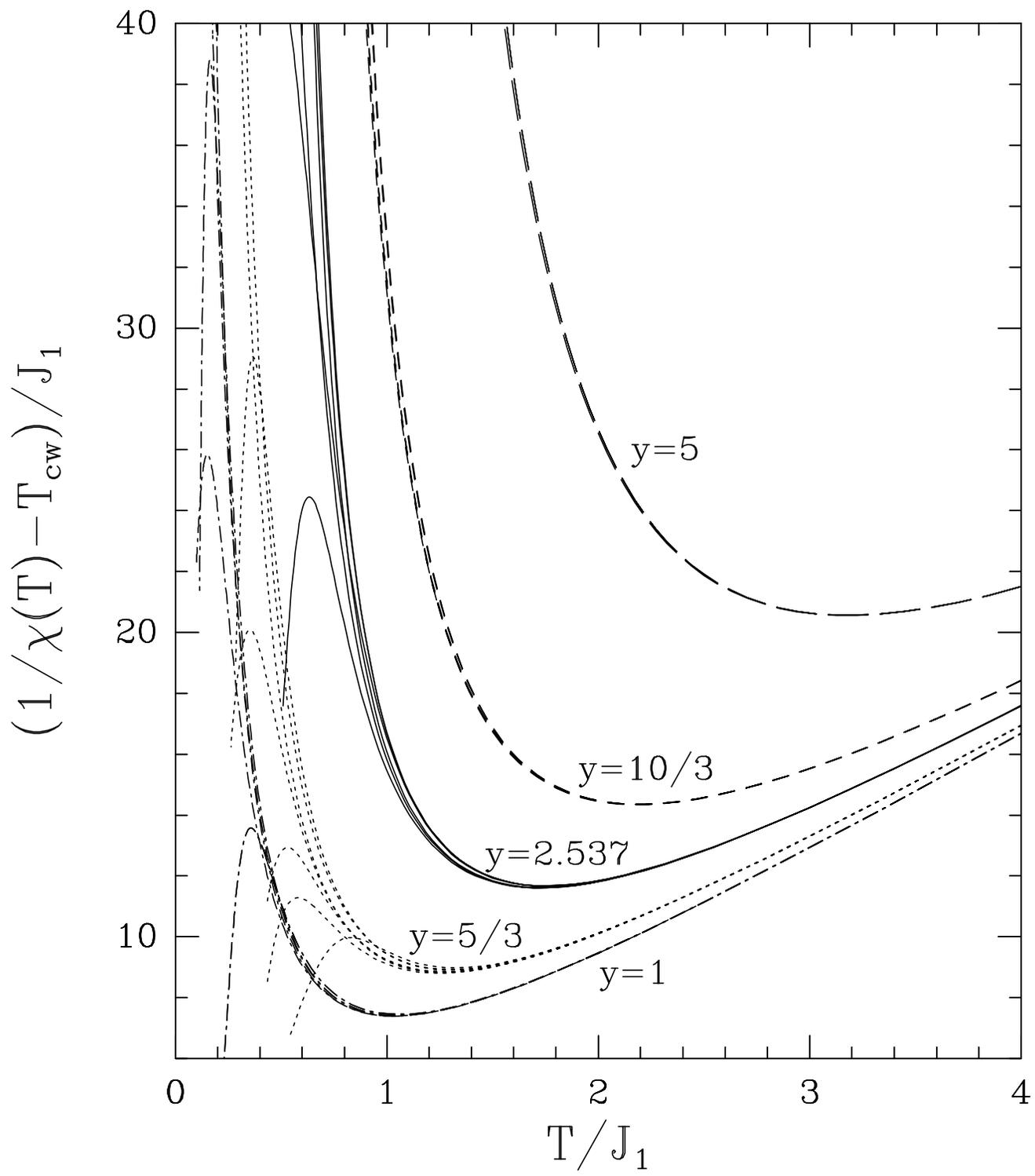
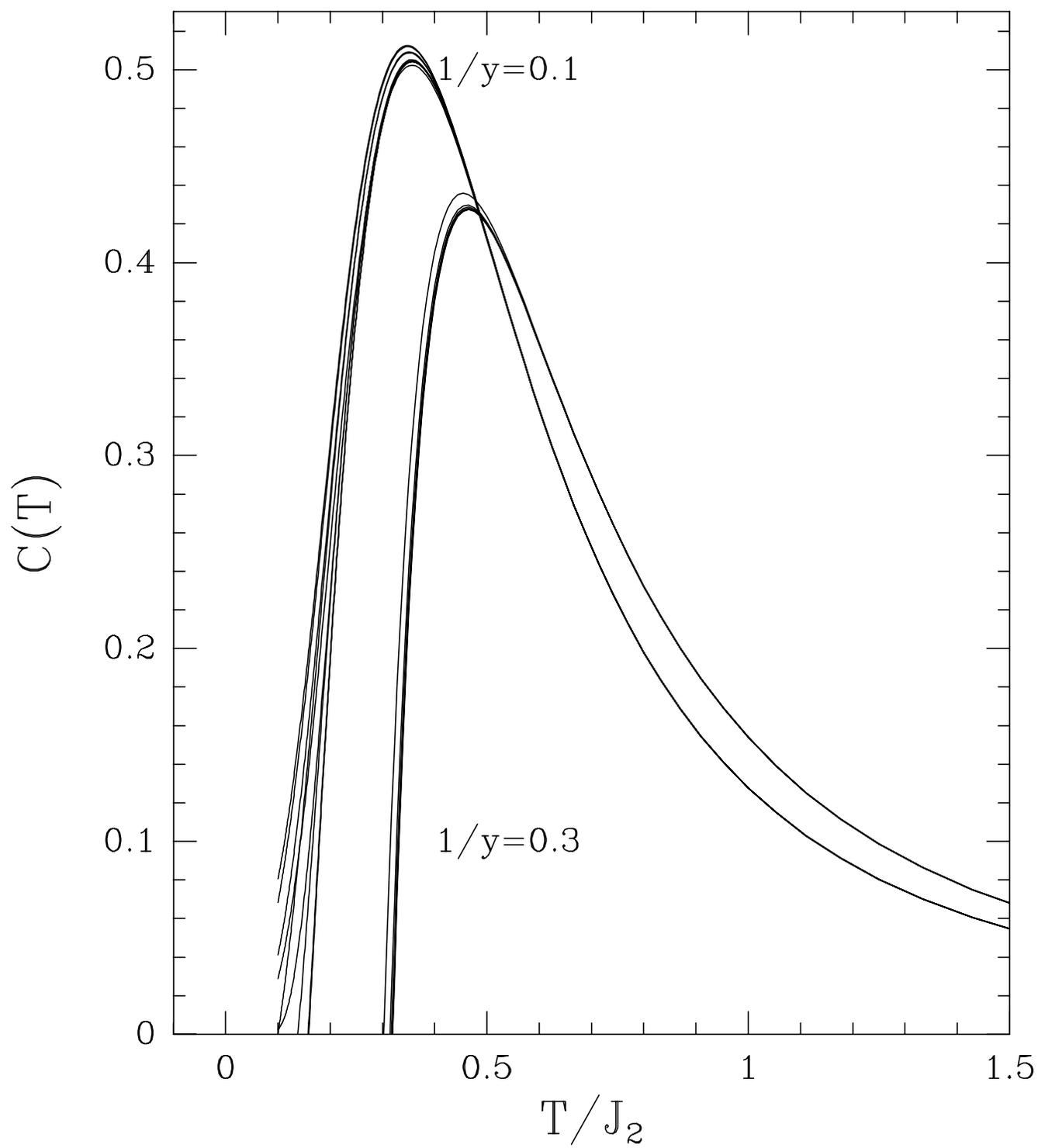
\end{document}